\documentclass[fleqn,twoside]{article}
\usepackage{espcrc2}

\usepackage{graphicx}
\usepackage[figuresright]{rotating}

\newcommand{\AmS}{{\protect\the\textfont2
  A\kern-.1667em\lower.5ex\hbox{M}\kern-.125emS}}

\title{
\vskip -110pt
{\large  
\mbox{} \hfill BNL-NT-04/32\\
}
\vskip 45pt
QCD Thermodynamics on lattice
       }

\author{
 P. Petreczky
\address[BNL]{Physics Department and RIKEN-BNL, Brookhaven National Laboratory,
        Upton, NY 11973 USA}                             
        }
       
\begin{document}

\begin{abstract}
I discuss recent developments in lattice QCD
thermodynamics on the nature of the transition
at finite temperature and density, equation of state,
screening of static charges and
meson spectral functions at high temperatures.
\end{abstract}

\maketitle
\section{Introduction}
It is expected that strongly interacting matter shows qualitatively
new behavior at temperatures and/or densities which are 
comparable or larger than the typical hadronic scale.
It has been argued by Hagedorn that production of hadronic
resonances may lead to a limiting temperature above which
hadronic matter can no longer exist \cite{hagedorn}.
The corresponding temperature can
be estimated to be $174MeV$ \cite{chapline74}.
On the basis of asymptotic freedom
one would expect that dominant degrees of freedom at very high
temperature and/or densities are quarks and gluons which are
no longer subject to confinement \cite{collins75}.
Such qualitatively different behavior of usual hadronic matter
with confinement and chiral symmetry breaking and this new
state of matter (usually called Quark Gluon Plasma) would 
suggest a phase transition at some temperature.
The existence of such phase transition was first shown in
the strong coupling limit of QCD \cite{firstlat},
followed by numerical 
Monte-Carlo studies of lattice SU(2) gauge theory which confirmed it
\cite{firstlat1}.

Since these pioneering studies QCD at finite temperature
became quite a large subfield of lattice QCD (for recent
reviews on the subject see Refs. \cite{frithjof,edwin,katz}). One of the 
obvious reasons for this is 
that phase transitions can be studied only
non-perturbatively. But even at high temperatures the physics
is non-perturbative beyond the length scales $1/(g^2~T)$
($g^2(T)$ being the gauge coupling) \cite{linde}. Therefore lattice QCD
remains the only tool for theoretical understanding of the
properties of strongly interacting matter under extreme condition
which is important for
the physics of the early universe and heavy ion collisions.
In this review I will discuss recent developments on studies of the
QCD phase transition at finite temperature and density, equation of state,
the problem of in-medium modifications of inter-quark forces
and meson spectral functions.

\section{QCD transition at finite temperature}

One of the basic questions we are interested in is what is
the nature of the transition to the new phase or if there is one 
at all, what is the temperature where it happens
\footnote{I will talk about QCD finite temperature transition
irrespective whether it is a true phase transition or a crossover
and $T_c$ will always refer to the corresponding temperature.}
 and what is 
the equation of state. In addition one may wonder what drives
the transition in QCD, which I will discuss in section 5.

In the case of QCD without dynamical quarks, i.e. SU(3) gauge theory
most of these questions have been answered. It is well
established that the phase transition is 1st order \cite{fukugita89}. 
Using standard and improved actions the corresponding transition
temperature was estimated to be $T_c/\sqrt{\sigma}=0.632(2)$ or
$265(1)MeV$ \cite{edwin} 
(assuming $\sigma=(420MeV)^2$ for the string tension).
RG-improved actions gave similar estimates for $T_c/\sqrt{\sigma}$
\cite{okamoto99}. 
When $T_c$ is expressed in units of the Sommer scale $r_0$
comparison between different action can be done in a quite
transparent way and gives  $r_0 T_c=0.7498(50)$ \cite{necco04}.
The equation of state of SU(3) gauge theory were also extrapolated 
to the continuum limit \cite{namekawa01,boyd96}.

The situation for QCD with dynamical quarks is much more difficult.
Not only because the inclusion of dynamical quarks increases the 
computational costs by at least two orders of magnitude but also
because the transition is very sensitive to the quark masses.
Conventional lattice fermion formulations break global 
symmetries of continuum QCD (e.g. staggered fermion
violate the flavor symmetry)  
which also introduces additional complications.

\begin{figure}
\includegraphics[width=2.5in]{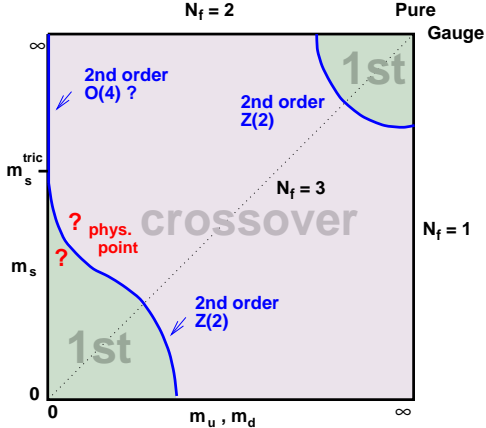}
\vspace*{-1cm}
\caption{The QCD phase diagram on the plane of $m_{u,d}$ and $m_s$.}
\label{columbia_plot}
\vspace*{-1cm}
\end{figure}
Since simulations at the physical value of the quark masses are
extremely expensive one usually studies the QCD phase 
diagram on the mass plane of u,d and s-quarks. This is 
schematically shown in Fig. \ref{columbia_plot}.
As mentioned before for infinitely heavy quarks
there is a deconfinement transition 
which is well known to be 1st order.
The 1st order nature of the transition persists for finite
values of the quark masses and terminates with a critical line
which belongs to $Z(2)$ universality class at quark masses 
corresponding to pseudo-scalar meson mass of 
about $2.5GeV$ \cite{heavyend}.
For the case of three flavors of light quarks universality arguments
suggest a 1st order chiral transition which was indeed observed
in lattice simulations with both standard and improved (p4) staggered 
fermion actions \cite{karsch03end,christ03}. 
The region of the 1st order transition ends at 
some critical line which belongs to $Z(2)$ universality class 
\cite{karsch03end,christ03}.
Note, however, that standard and improved fermions give
different estimates for the value of quark (pion) mass where the 1st
order region ends, $m_{\pi}^c \simeq 270MeV$ for standard action
\cite{christ03} and $m_{\pi}^c=67(18)MeV$ for p4 action \cite{karsch03end}.
The transition temperature in the chiral limit
for $N_t=4$ was estimated to be $T_c=154(8)$ \cite{karsch01nfdep}.

For two massless flavors one expects a 2nd order transition
with $O(4)$ universality class. Simulations with staggered fermions
(both standard and improved) did not confirm this expectations.
On contrary a recent analysis with standard staggered action on large
lattices suggests that transition maybe weakly 1st order 
\cite{delia04}. On the other
hand simulations with Wilson fermions show finite size scaling consistent
with $O(4)$ exponents \cite{alikhan01}. 
Despite these disagreements the chiral critical
temperature for two flavor QCD turns out to be consistent for Wilson
and staggered formulations: for
Wilson fermions one has $T_c=171(4)MeV$ \cite{alikhan01}, 
while for improved staggered (p4) $T_c=173(7) MeV$ \cite{karsch01nfdep}
( both refer to $N_t=4$ lattice).
First attempts to study the finite temperature transition in 
two flavor QCD
with chiral fermions which have all the symmetries of continuum
QCD have been done in Ref. \cite{fodor04overlap}.

The most interesting question is about the nature of the transition
in the physical case of two light and one heavier quarks and where the
physical point in Fig. \ref{columbia_plot} is located. Simulations for
this case have been performed  using standard \cite{aoki99,fodor04} 
and improved staggered formulations 
\cite{milc04thermo,milc_lat04,milc_lat03,milc_lat02}. 
The most recent calculations with standard staggered fermions
were performed on $N_t=4$ lattices 
and quark 
masses very close to physical ones corresponding 
to the lightest pion mass of about $150MeV$ ( $m_{\pi}/m_{\rho}=0.188(2)$).
Calculations with improved staggered (Asqtad) action 
were done using three different
lattice spacings ($N_t$) and several quark masses, with lightest quark mass
being only 2.5 times heavier than the physical value. Both 
studies have found no evidence for a true phase transition but only
a crossover. The crossover temperature was estimated to be $T_c=164(2)MeV$
(the errors is only statistical and $r_0=0.5fm$ was assumed) 
for standard staggered action \cite{fodor04}  
and $169(12)(4)MeV$ for Asqtad \cite{milc_lat04}. The later is
an extrapolated value. The continuum and chiral extrapolation was done with
$T_c(N_t)=T_c+c_1 (m_{\pi}/m_{\rho})^d+c_2/N_t^2$ Ansatz with $d=1.08$
motivated by $O(4)$ scaling. 
The coefficient $c_2$ turns out to be 
quite small as expected; in Asqtad action the anomalously large 
$\alpha_s a^2$ corrections are removed. The second error is the estimate
of $T_c$ is the uncertainty in the scale (lattice spacing) \cite{milc_lat04}. 
If expressed in units of the Sommer scale $r_0$ 
both standard action and Asqtad action give a consistent estimate
for $T_c$. When using the 
most recent value $r_0=0.467fm$ \cite{milc04} 
and rescaling the results one gets the value
$T_c=176(2) MeV$ for the standard action.

\begin{figure*}
\includegraphics[width=2in]{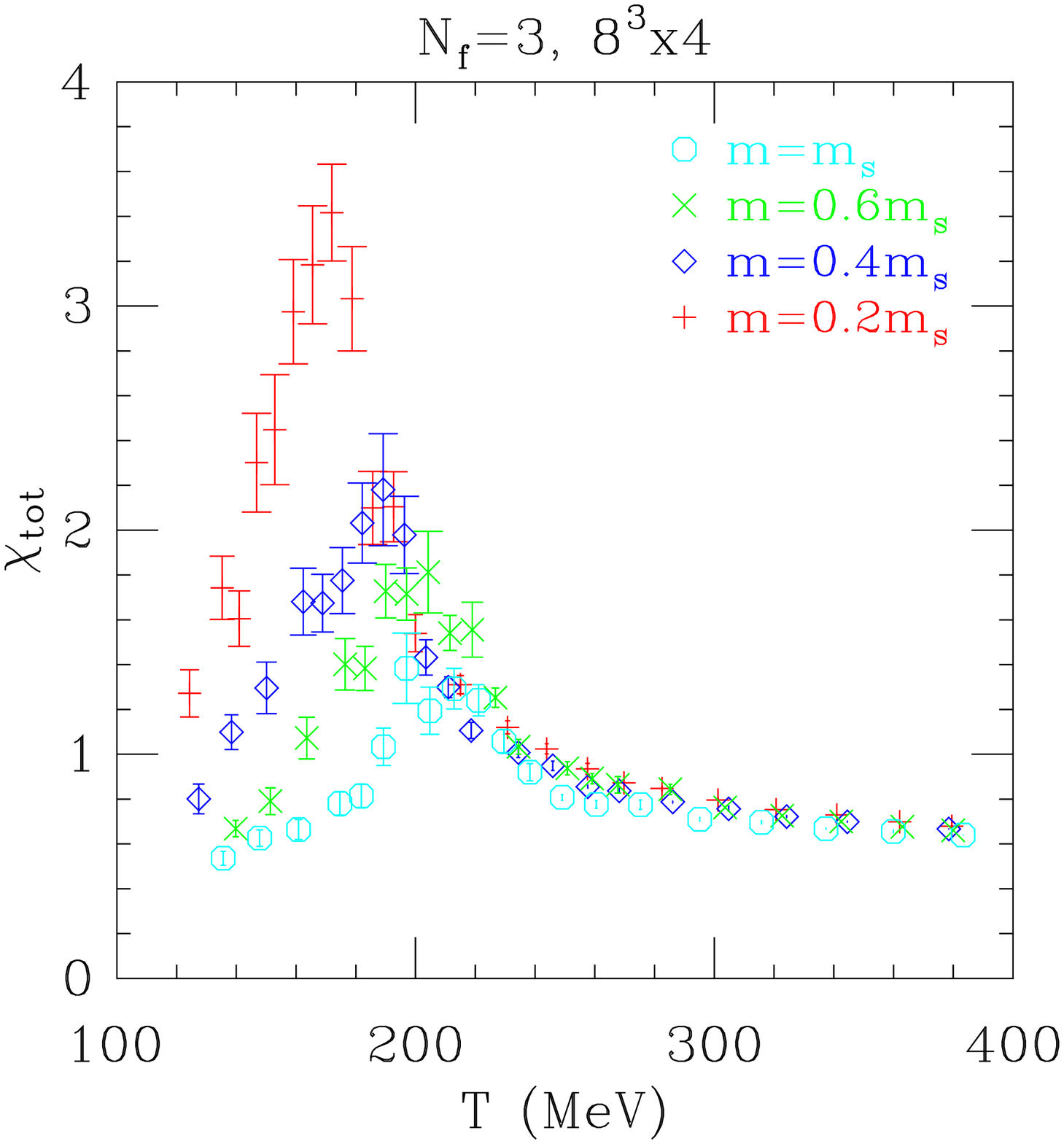}
\includegraphics[width=2in]{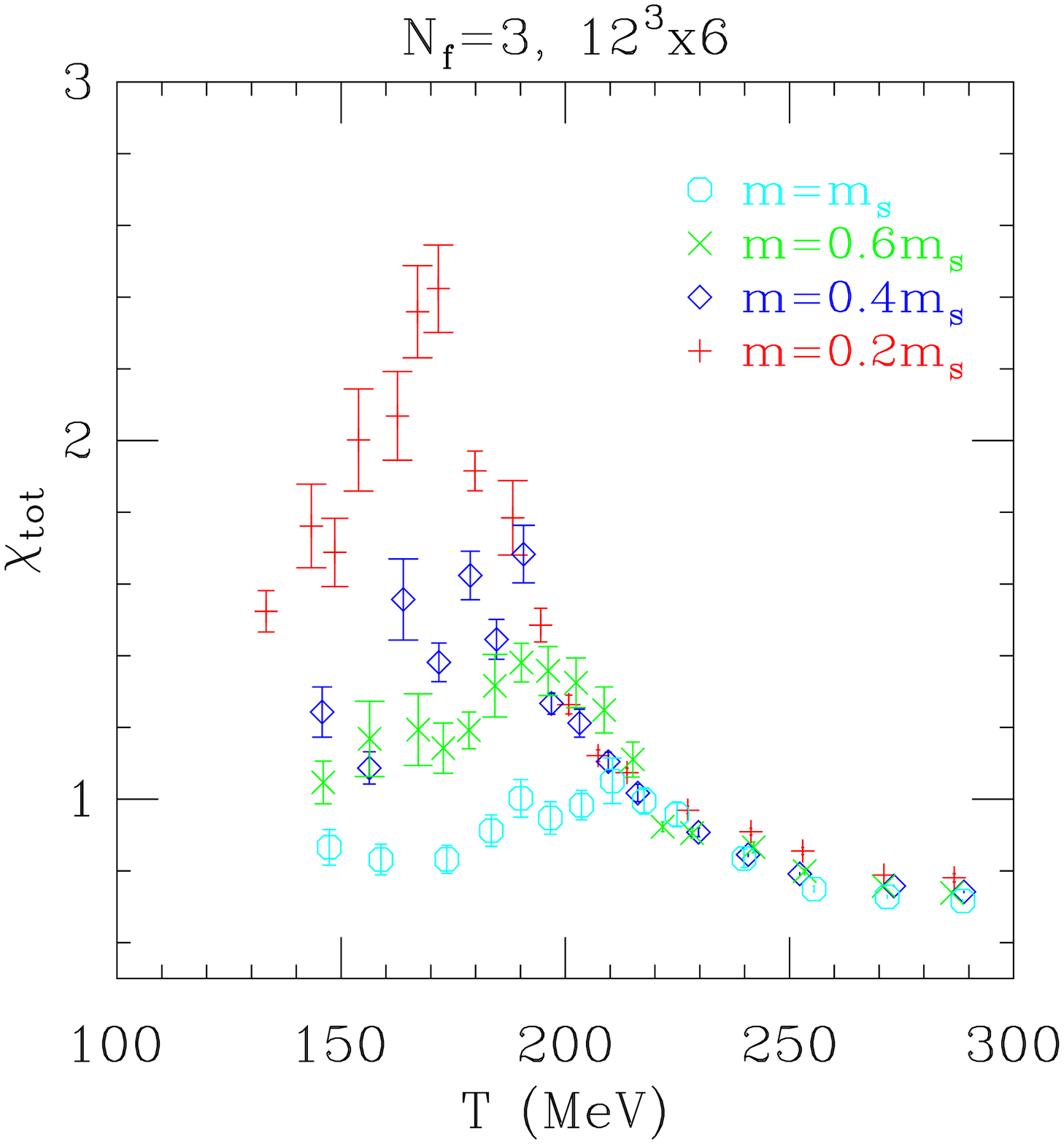}
\includegraphics[width=2in]{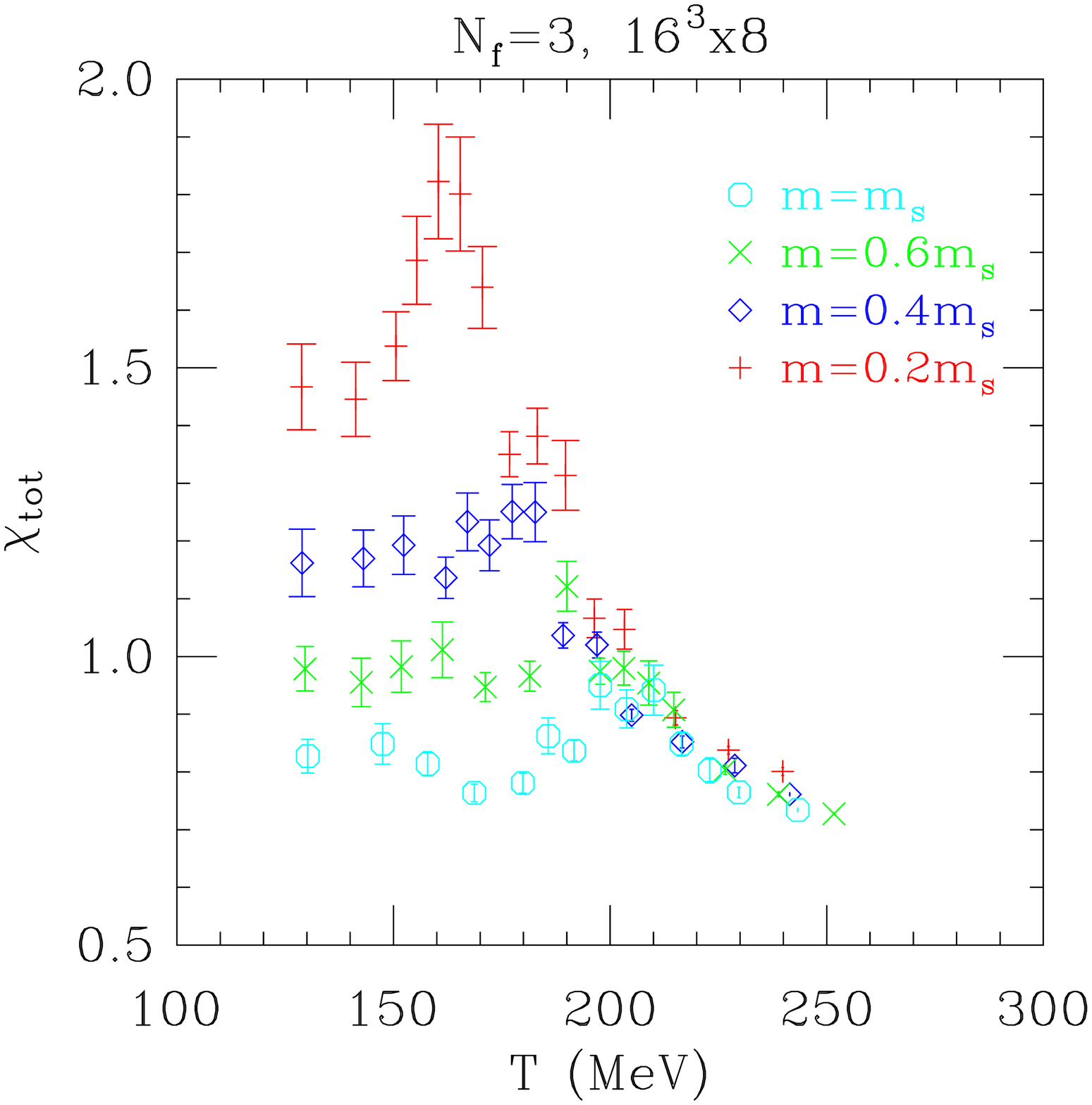}
\vspace*{-2cm}
\caption{The chiral susceptibility in three flavor QCD
for $N_t=4,~6$ and $8$.}
\label{milc_chipp_nf3}
\vspace*{-0.5cm}
\end{figure*}

It seems that different fermion actions give very similar estimate
for the transition temperature, which interestingly enough appears
to be very close to the old prediction from the resonance gas model.
Why are there so large discrepancies concerning the order of
the phase transition? As mentioned before staggered fermions violate
the flavor symmetry. This results in the fact that there is only one
Goldstone pion, the other pions remain massive in the chiral limit for
any finite lattice spacing. For typical lattice spacings used in thermodynamic
studies ($a=0.3-0.2 fm$) the splitting between Goldstone and non-Goldstone
pions can be quite large, so the lightest non-Goldstone pion turns out to
be heavier than the kaon. Clearly it is important to improve the flavor
symmetry of staggered fermions when one wants to make statement about 
the nature of the transition at the physical point. This is usually done
by introducing fat links in the fermion hopping term 
(see e.g. Refs. \cite{karsch01nfdep,milc04thermo}) . Typically the fat
links which are used are the sum of a single link and staples and
are not elements of the SU(3) group. In staggered HYP action the fat links
are elements of SU(3) gauge group \cite{hasenfratz01}. 
It has been shown that projecting the
fat links onto SU(3) group substantially improves the flavor 
symmetry  \cite{hasenfratz01}.
Finite temperature simulations with HYP actions gave a number of 
interesting results \cite{hasenfratz01}. 
For QCD with four light flavors using standard staggered action the transition
was found to be first order in accord with universality arguments
\cite{1stnf4,newnf4}. The only problem is that the critical quark
mass where the transition turns into crossover was found to be strongly
dependent on the lattice spacing \cite{newnf4}, 
it decreases by more than a factor of three when going from $N_t=4$
to $N_t=6$. In Ref. \cite{hasenfratz_nf4} it was found that introducing
unprojected fat links make the transition weaker, while with the use of
HYP smeared fat links the first order transition is washed out completely.
In fact for HYP action, no first order transition was observed  
even for very small quark masses both for $N_t=4$ and $N_t=6$ 
lattices \cite{hasenfratz_nf4}.
There are also simulations with HYP action at $N_t=4$
in $2$ flavor as well as at $N_t=4$ and $6$ in
$2+1$ flavor QCD which indicate that the crossover is much smoother
than for standard staggered action for the same quark mass and same
lattice spacing \cite{hasenfratz_nf2,fodor_hyp}. 
Can improvement of flavor symmetry in the staggered 
fermion action smoothen the 
transition and eventually wash out any phase transition for the quark masses
where standard action predict a phase transition?
In this respect it is interesting to study the smoothness (sharpness)
of the crossover as function of the lattice spacing. Typically
susceptibilities of order parameter like quantities, e.g. chiral condensate
$\bar \psi \psi$, show a pronounced peak for a rapid crossover. 
For a true phase
transition the height of the peak diverges in the thermodynamic limit, while
for a crossover it approaches a finite value. 
The MILC collaboration has calculated the susceptibility of the chiral 
condensate for three flavor QCD at three different lattice spacings 
\cite{milc04thermo,milc_lat04,milc_lat03,milc_lat02}. 
The results are shown in Fig. \ref{milc_chipp_nf3}. 
One can see a pronounced
peak for $N_t=4$, however, the peak height relative to
the low temperature value of the chiral susceptibility 
decreases with increasing 
$N_t$. This means that as the flavor symmetry gets restored by going to 
finer lattices (smaller $N_t$) the crossover becomes smoother.
Note that the chiral susceptibility in Fig. \ref{milc_chipp_nf3}
is given in lattice units. To convert to physical units, say units of
$T^2$ the data should be multiplied by $N_t^2$ which will make 
the chiral susceptibility look very different for different
$N_t$.
Thus it is likely that the first order transition
observed so far with staggered action is just a lattice 
artifact.

\section{Phase diagram at $\mu>0$}
For a long time there was no considerable progress in studying the phase 
diagram of QCD at finite chemical potential. This is because the 
complex fermion determinant makes importance sampling based
simulations impossible. The multi-parameter re-weighting method by 
Fodor and Katz was a partial breakthrough which allowed to study the 
phase diagram of QCD also at finite chemical potential 
\cite{fodor01st,fodor01}.
Soon after
also other methods based on Taylor expansion around $\mu=0$ 
\cite{alton02,alton03} and
analytic continuation to imaginary chemical potential 
\cite{philipsen02,philipsen03,delia03} were suggested.
All three methods were discussed in detail by S. Katz at Lattice 2003.
Therefore I will emphasize mostly the most recent results leaving technical
details aside.

\begin{figure}
\vspace*{-0.8cm}
\includegraphics[width=2.3in]{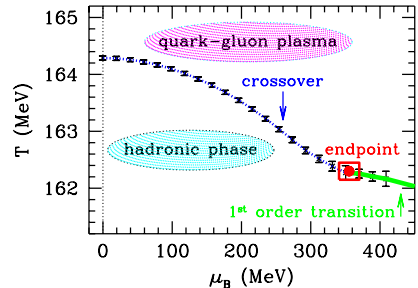}
\vspace*{-2cm}
\caption{The phase diagram at finite chemical potential obtained
using multi-parameter re-weighting \cite{fodor04}. }
\label{fodor_new}
\vspace*{-0.7cm}
\end{figure}

In the previous section we have seen that all available lattice
data suggest that there is no phase transition in real QCD but only a 
smooth crossover. In this case a very interesting phase diagram in the
plane of temperature and chemical potential $(T,\mu)$ has been
conjectured.
At large chemical potential a 1st order transition is
expected and thus if there is a crossover at $\mu=0$ the phase
transition line should end at some critical point $T_E,\mu_E$.
The important question is whether we can locate this end-point on
the lattice. The first attempt has been made by Fodor and Katz using
multi-parameter re-weighting \cite{fodor01}. 
At each value of $\mu$ the pseudo-critical
point in the gauge coupling $\beta$ was located using Lee-Yang zeros,
zeros of the partition function for complex $\beta$ which are closest 
to the real axis.
In the crossover region Lee-Yang zeros have non-vanishing imaginary parts
in the thermodynamic limit. For a true phase transition, on the other hand,
the imaginary part of Lee-Yang zeroes vanishes in the infinite volume limit.
Thus inspecting the finite size scaling of the imaginary part of
Lee Yang zeros one can determine whether there is a phase transition and
eventually determine its order; the real part of Lee-Yang zero gives the
location of the critical $\beta$ (or equivalently the temperature).
Using this procedure with standard staggered action, quark masses   
about three times larger than the physical one and lattices 
$4^4,~6^3 \times 4,~8^3 \times 4$ they estimated the location of the
critical end-point at $T^E=172(3)MeV,~\mu_B^E=725(35)MeV$ \cite{fodor01}
(note that the baryon chemical potential $\mu_B=3 \mu$ with $\mu$
being the quark chemical potential). 
Very recently
they extended their studies by including larger lattice volumes and 
values of the quark masses very close to the physical ones \cite{fodor04}.
The result of their new analysis is summarized in Fig. \ref{fodor_new}. 
The endpoint was found to be located
located at $T^E=162(2)MeV, \mu_B^E=360(40)MeV$. It is evident that the
end-point is very sensitive to quark mass. Calculations using the 
method of analytic continuation in 2+1 flavor QCD \cite{philipsen02}
with the same 
quark masses as used in Ref.\cite{fodor01} 
give results which are in very good
agreement with findings of Fodor and Katz for the position of the
pseudo-critical line. There are also calculations based on Taylor
expansion  in 2 flavor QCD with
somewhat heavier quark masses which gives qualitatively similar results.
In the case of 3 flavor QCD using re-weighting with Taylor 
expansion the critical endpoint was estimated to be 
$\mu_B^E=120(30)MeV$ for $m_{\pi} \simeq 140MeV$ \cite{karsch03end}. 

\section{Equation of state}
Lattice calculations of the equation of state are difficult
for at least two reasons. First, the calculation of the energy density and
pressure needs subtraction of the zero temperature (vacuum) part.
This makes computational expenses grow with lattice spacing 
as $a^{-11}$ compared to $a^{-7}$ growth for other quantities.
The second reason is that the equation of state more than any other
quantity is sensitive to high momentum modes and therefore to
cut-off effects. Indeed calculations with standard staggered
action as well as Wilson (clover) action show large dependence on
the lattice spacing ($N_t$) \cite{milc_eos,alikhan_eos}. 
Using improved staggered fermions cut-off effects in thermodynamic
quantities can be greatly reduced \cite{heller99}.
In Ref. \cite{karsch_eos}
pressure and energy density have been calculated with p4 action
for $N_t=4$ and different number of flavors and for quark masses
corresponding to pseudo-scalar (pion) mass of about $700MeV$.
When expressed in units of $T^4$
both pressure and energy density turn out to be much larger than in
SU(3) gauge theory reflecting the presence of additional 
degrees of freedom (quarks). An interesting question is what is
the energy density at the pseudo-critical point. Calculations
with both Wilson and p4 staggered actions yield consistent estimates,
$\epsilon_c=(6 \pm 2)T_c^4$ \cite{karsch01qm},
which is in turn not too sensitive to the quark mass and flavor. 
Assuming that this holds down to physical quark masses 
and taking the most recent value $T_c=169MeV$ one can
estimate the (pseudo)critical energy density to be $(0.6 \pm 0.3)GeV/fm^3$. 

\begin{figure}
\includegraphics[width=2.1in]{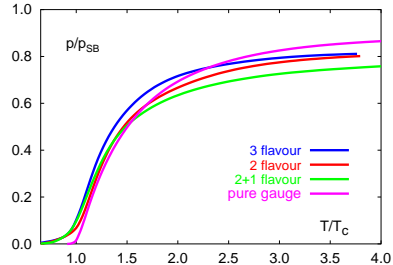}
\vspace*{-1cm}
\caption{The pressure normalized by Stefan-Boltzmann limit
for different number of flavors.}
\vspace*{-0.7cm}
\label{ppsb}
\end{figure}

In Fig. \ref{ppsb} 
we show the pressure normalized
by continuum Stefan-Boltzmann limit for different number of quark flavors.
One can see almost no flavor dependence in $p/p_{SB}$ for $T>T_c$,
i.e. the flavor dependence of the pressure in the high temperature
phase can be well described by flavor dependence of $p_{SB}$.
Note that the pure gauge result was extrapolated to the continuum
limit. This means that if the approximate flavor independence is true
the p4 action results at $N_t=4$ are already close to the continuum limit.
Nonetheless one should keep in mind that the above calculations
were done at fixed $m_{quark}/T$ and not along the lines of constant physics.
It has been shown in Ref. \cite{csikor04} 
that keeping $m_{quark}/T$ fixed instead of keeping
the quark mass $m_{quark}$ constant in physical units can reduce the pressure
by $9\%$ at $2.5T_c$.

Recently the 
pressure has also been calculated at non-zero chemical potential
using both the Taylor expansion technique \cite{alton03} 
and multi-parameter re-weighting \cite{csikor04}.
More precisely one calculates the baryonic contribution to the
pressure $\Delta p(T,\mu)=p(T,\mu)-p(T,\mu=0)$.
The ratio  $\Delta p/\Delta p_{SB}$, 
where $\Delta p_{SB}$ is the ideal gas limit of $\Delta p$,
is almost independent of the chemical
potential \cite{alton03,csikor04}.
This  means that the dependence of the pressure on the chemical
potential is well described by that of ideal quark-gluon plasma.
Taylor expansion method with p4 action gives similar result \cite{alton03}
which will be discussed in the next section. Here we only note that 
despite the similar behavior of $\Delta p/\Delta p_{SB}$ 
in the two methods there is
a problem of cut-off dependence, in. Ref. \cite{csikor04} $\Delta p_{SB}$
is calculated for $N_t=4$  lattice , while in Ref. \cite{alton03}
the continuum value for $\Delta p_{SB}$ was used. 

\section{Comparison with the resonance gas model}

The QCD transition has two aspects : deconfinement and
chiral symmetry restoration. In absence of quarks in the fundamental
representation an order parameter (the expectation value of
the Polyakov loop) and a symmetry ($Z(N)$) can be linked to
the deconfinement transition. In the more general case it is more
meaningful to define the deconfinement transition as rapid
increase of degrees of freedom. The chiral symmetry restoration is linked
to the behavior of the chiral condensate and lightest hadrons 
with increasing temperature. The interconnection between chiral
and deconfinement aspects of the QCD transition is subject to intensive 
discussion in the literature \cite{interconn}. 
There are at least two question which
can be addressed in this respect. Does the mass of the lightest
excitation (e.g. pion) determines the value of the transition temperature?
What drives the QCD transition?
The answer to the first question is probably ``no''  and can be explained
using several reasonings. First, if the mass of the lightest
excitation determines the value of the transition temperature
it would be difficult to understand why the transition temperature in
the SU(3) gauge theory, where the lowest excitation is the scalar
glueball of mass of about $1.5GeV$ is only $50\%$ larger than the
transition temperature in full QCD where pions of mass $140MeV$ are the
lowest excitations. Second in terms of functional integral the finite
temperature system can be viewed as a zero temperature system in
a slab of extension $1/T$ ($T$ being the temperature). The non-perturbative
features of the vacuum (causing confinement and chiral symmetry breaking)
are obviously lost when the size of the slab becomes too small. There is
no obvious connection between the ``critical'' size of the slab where
the relevant non-perturbative features are lost and the mass of the lowest
excitation \footnote{I thank J. Kuti for bringing up this point.}.

In the following we are going to discuss the resonance gas model
which provides an answer to both questions raised above in terms of
physical degrees of freedom. Despite being a model it can explain many
features of the available lattice data and can provide a standard reference for
bulk thermodynamic properties in the low temperature phase similarly to the 
ideal quark-gluon plasma case ( Stefan-Boltzmann limit) which serves as a reference 
in the high temperature phase.

\begin{figure}
\vspace*{-0.5cm}
\rotatebox{-90}{\includegraphics[width=1.7in]{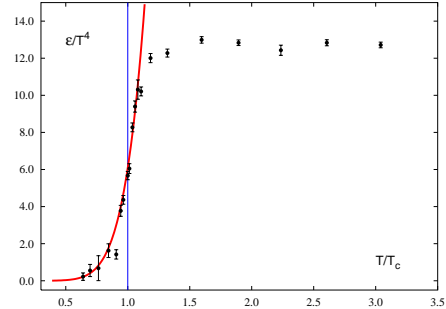}}\vspace*{-0.5cm}
\caption{The energy density in resonance gas model (thick line)
against lattice data with p4 improved staggered fermions \cite{karsch_rg}.}
\vspace*{-0.7cm}
\label{resonance_eps}
\end{figure}

In the resonance gas model the partition
function is a sum of one- particle partition functions of all
mesons, baryons and their resonances \cite{karsch_rg}. 
In Fig. \ref{resonance_eps}
I show the lattice data for the energy density 
(obtained with $p4$ action) in comparison with the resonance gas model
which includes about $2000$ degrees of freedom  corresponding
to resonances with  masses below $2GeV$. 
Since lattice calculations are done at unphysical quark masses the
masses of experimentally know resonances have been extrapolated 
upward in quark mass to take this into account \cite{karsch_rg}. 
Note that pions contribute
only $14\%$ to the energy density at the transition 
temperature \cite{karsch_rg}. 
As one can see the resonance gas model can described the 
lattice data very well till temperatures of about $1.1T_c$.

\begin{figure}
\vspace*{-0.5cm}
\includegraphics[width=1.7in]{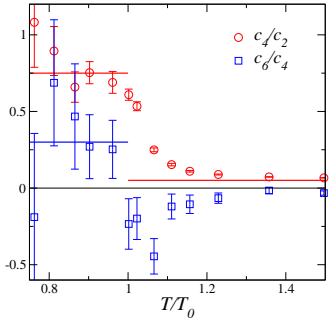}
\vspace*{-1cm}
\caption{
The ratios of the expansion coefficients $c_n$
calculated in resonance gas model \cite{karsch_rg1}
and on the lattice \cite{ejiri04}.
}
\vspace*{-0.8cm}
\label{ratios_cn}
\end{figure}

One immediate consequence of
the resonance gas model is the factorization of the baryonic
pressure

\begin{equation}
\frac{\Delta p}{T}=\frac{p(T,\mu)-p(T,\mu=0)}{T}=
F(T) \cosh(\frac{3 \mu}{T}),
\label{deltap_reson}
\end{equation}
which holds as long $(m_{nucleon}-3 \mu) > T$ \cite{karsch_rg1}.
On the lattice one can calculate the pressure at finite chemical
potential in form of the Taylor expansion
\begin{equation}
\frac{\Delta p}{T}=c_2(T) (\frac{\mu}{T})^2 + c_4(T) (\frac{\mu}{T})^4+
c_6(T) (\frac{\mu}{T})^6 + ...
\label{deltap_taylor}
\end{equation}
The factorization formula has important consequences for the 
baryonic pressure and its derivatives. The ratio of the baryonic
pressure and its derivatives with respect to
$\mu$ (quark number and quark number susceptibility)  
are function of $\mu/T$ only \cite{karsch_rg1}. 
The ratios of the expansion coefficients
in Eq. (\ref{deltap_taylor}) are those of $\cosh$: 
$c_4/c_2=3/4$, $c_6/c_4=0.3$. These predictions can be compared 
with lattice data \cite{ejiri04} as shown in Fig. \ref{ratios_cn}.
The figure clearly shows that the resonance gas model works quite well at low
temperatures. 

\begin{figure}
\vspace*{-0.2cm}
\includegraphics[width=2.2in]{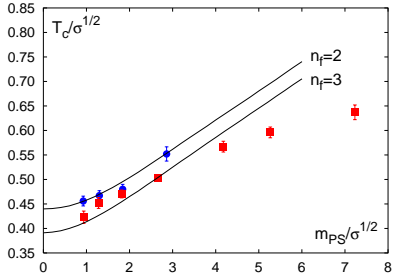}
\vspace*{-1cm}
\caption{
The dependence of the deconfinement temperature on
the pseudo-scalar mass calculated from lattice 
simulations \cite{karsch01nfdep} and resonance gas model \cite{karsch_rg}.
}
\vspace*{-0.6cm}
\label{tc_reson}
\end{figure}

Now we can return to the question what drives the transition in QCD.
The energy density at the pseudo-critical point can be estimated
to be $\epsilon_c \simeq 0.6GeV/fm^3$ (see previous section). 
Assuming that this critical energy
density is independent of the quark mass and number of flavors,
as suggested by the existing lattice data,
one can study the mass and flavor dependence of the transition
temperature $T_c$. This is shown in Fig. \ref{tc_reson} which shows
that the resonance gas at fixed energy density of $0.8GeV/fm^3$
seems to explain the quark (pion) mass dependence of $T_c$ as 
long as $m_{PS}/\sqrt{\sigma}<4$. 
For larger pseudo-scalar masses
the discrepancies can be explained by the fact that at such large 
quark mass glueball states become important. It has been shown that
inclusion of glueball states with thermal mass reduction can account
for this discrepancy \cite{karsch_rg}. 

Thus we see that deconfinement happens once
the energy density reaches a value of about $0.8GeV/fm^3$ and 
does not depend strongly on the details of the spectrum but only
on its exponential rise at large resonance masses. The fact that
space gets (over)populated by hadron resonances near $T_c$ also means
that there is little space left for the non-perturbative vacuum if
one assumes a simple MIT-bag motivated picture where the vacuum inside
a hadron is perturbative. This may lead to the decrease of the average
value of chiral condensate. Also production of resonances increases
the entropy, and the Polyakov loop, which can be related to the 
additional entropy generated by a static meson 
(see next section, for entropy increase due to adding an extra baryon into
the system see \cite{kratochvila04lat} ), therefore rapidly increases near $T_c$.

\section{The free energy of static quarks}

It is customary to study in-medium modification of inter-quark forces, e.g.
screening in terms of the free energy of static quarks separated by some
distance $r$. Following McLerran and Svetitsky the partition
function of the system with static quark anti-quark pair ($Q \bar Q$)
normalized by the partition function of the system without static
charges can be expressed as correlation function of temporal Wilson 
lines \cite{larry}
\begin{equation}
\frac{Z_{Q\bar Q}(r,T)}{Z(T)}=\langle W(\vec{r}) W^{\dagger}(0) \rangle,
\end{equation}
\begin{equation}
W(\vec{x})={\cal P} \exp(-\int_0^{1/T} d\tau A_0(\vec{x},\tau)).
\end{equation}
${\rm Tr} W$ is usually referred to as Polyakov loop. Taking the logarithm
of the above expression gives the free energy difference 
of the system with static quarks and without them at fixed temperature
$T$. Since the quark and anti-quark can be either in color singlet
or color octet state one should separate those contributions using
the appropriate projection operators which gives the following
expressions for the singlet and octet free energies \cite{brown79,nadkarni86}
\begin{eqnarray}
&&
\exp(-F_1(r,T)+C)=
\frac{1}{3} \langle {\rm Tr} W(\vec{r}) W^{\dagger} (0) \rangle ,\\
&&
\exp(-F_8(r,T)+C) = -\frac{1}{24}
\langle {\rm Tr} W(\vec{r}) W^{\dagger} (0) \rangle \nonumber\\
&&
+\frac{1}{8}  \langle {\rm Tr} W(\vec{r}) {\rm Tr} W^{\dagger} (0) \rangle .
\label{f1f8_def}
\end{eqnarray}
One can define also the color average free energy which is the thermal
averaged of the free energies in color singlet and octet channels 
\cite{larry,nadkarni86}
\begin{eqnarray}
&&
\exp(-F(r,T)+C) = \frac{1}{9} \langle 
{\rm Tr} W(\vec{r}) {\rm Tr} W^{\dagger} (0) \rangle \nonumber\\
&&
=\frac{1}{9} \exp(-F_1(r,T))+\frac{8}{9} \exp(-F_8(r,T)).
\end{eqnarray}
The normalization constant $C$ in the above equations will be
defined below.
The color averaged free energy is explicitly gauge invariant,
while the others need gauge fixing or the Wilson line to be
replaced by the gauge invariant Wilson line as this was done
in Ref. \cite{philipsen02octet}. 
The results to be presented here were obtained using the
Coulomb gauge as in this gauge a transfer matrix can be 
defined and the free energies have a meaningful zero temperature
limit. In fact it was shown that the Coulomb gauge and gauge
invariant definition of the singlet free energy give numerically
the same result \cite{philipsen02octet}.

\begin{figure}
\includegraphics[width=2.4in]{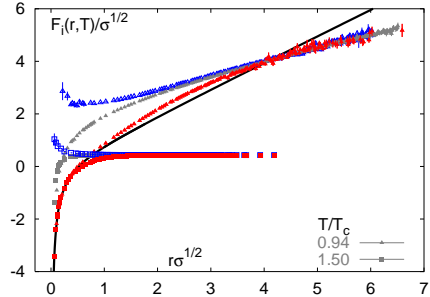}
\vspace*{-0.9cm}
\caption{The color singlet, octet and averaged free energy
in SU(3) gauge theory above and below 
deconfinement \cite{zantow03,kaczmarek03lat}.}
\label{free_all}
\vspace*{-0.5cm}
\end{figure}

In Fig. \ref{free_all} we show the singlet, octet and averaged 
free energies in $SU(3)$ gauge theory 
\cite{kaczmarek02,zantow03,kaczmarek03lat}. The normalization constant
$C$ was chosen such that the singlet free energies coincides with
the zero temperature potential at the shortest distance.
Below the deconfinement temperature all free energies are linearly rising,
indicating confinement. The string tension, however, is smaller than
in the vacuum. Above deconfinement all the free energies approach
the same constant value at asymptotic distances indicating the 
screening in plasma phase. Thus the free energies of static 
$Q \bar Q$ pair provide a true test of confinement. If this free energy
is finite at infinite separation 
the system is deconfined and otherwise it is confined.
Usually the expectation value of the Polyakov loop 
$\langle {\rm Tr} W \rangle$ is considered to be an order parameter
for deconfinement and it is said that it is related to the free
energy of a static quark $|\langle {\rm Tr} W \rangle| \sim \exp(-F_Q(T)/T)$.
This quantity, however, does not have a meaningful continuum limit; for
any temperature it vanishes in the continuum limit. One can define
a renormalized Polyakov loop as 
$L_{ren}(T)=\exp(-F_{\infty}(T)/(2T))$ with $F_{\infty}(T)$ being the 
asymptotic value of the $Q \bar Q$ free 
energy at infinite separation \cite{kaczmarek02}.
This provides an order parameter which is cutoff independent and
has a well defined continuum limit \cite{kaczmarek02}.
While in the deconfined phase it is not difficult to accept the 
concept of the octet free energy, what it means in the confined phase
is not quite clear. 
The construction in Eq. (\ref{f1f8_def}) fixes only the relative
color orientation of static charges, e.g. $(Q \bar Q)_8$ being
in octet state. In order for such object to exist in the confined 
phase the net color charge must be compensated by gluons. Such an
object can be identified with excited state of the string or
hybrid potential \cite{juge01,bali01}. 
In Ref. \cite{jahn04} the definitions of the singlet and octet
free energies were applied in the zero temperature limit. It was found
there that the definition of the octet free energy does not project
onto the string excitation as expected. This is because the amplitudes
$c_n(r)$ in the spectral decomposition of the correlators
$\sum_n c_n(r) \exp(-E_n (r)/T)$ have non-trivial $r$-dependence 
\cite{jahn04}.
At short distances where the $r$-dependence of $c_n$ is likely to  vanish
and the octet free energy in (\ref{f1f8_def}) could give the hybrid potential.

\begin{figure}
\includegraphics[width=2.7in]{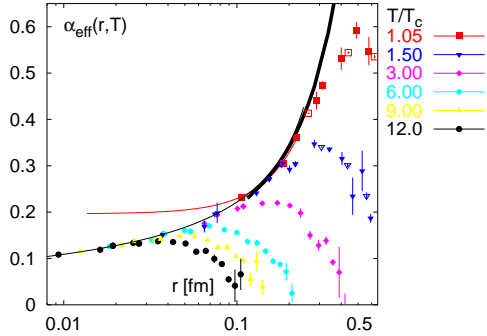}
\vspace*{-1cm}
\caption{The effective running coupling constant for 
different temperatures \cite{kaczmarek04}.}
\label{alpha_eff}
\vspace*{-0.8cm}
\end{figure}

Notice that at short distances the singlet free energy coincides 
with the zero temperature potential in a certain distance range and 
is temperature independent, reflecting the fact that medium effects
are not important at short distances. This fact motivated our
choice of the renormalization constant. 
To study the onset of medium effects and characterize the 
strength of interaction in the plasma it is convenient
to introduce the effective running coupling constant
$\alpha_{eff}(r,T)=\frac{3}{4} r^2 \frac{d F_1(r,T)}{dr}$ \cite{kaczmarek04}
which is shown in Fig. \ref{alpha_eff}.
At short distances the scale which determines the 
running of the coupling constant is given
by the distance $r$ and running follows the 3-loop beta function. 
At larger distances 
the fast running of the coupling constant mimics the presence of 
the linear term in Cornell parameterization of the zero temperature 
potential and therefore is of non-perturbative origin. 
It is interesting to note that this non-perturbative feature
of the running coupling constant survives also in the deconfined phase
as shown in Fig. \ref{alpha_eff}.
The effective
running coupling starts to deviate from the zero temperature result 
at some distance and eventually decreases indicating the onset of medium
effects and screening. The distance where it happens is obviously
temperature dependent. Note that at distances where screening sets in, the 
singlet free energy does not have a simple screened Coulomb form
expected in finite temperature perturbation theory. Only at somewhat larger
distances the singlet free energy can be fitted by the form 
$\frac{4}{3}\alpha_s(T) \exp(-m_1(T) r)/r$ 
motivated by perturbation theory \cite{kaczmarek04}.
This defines then a coupling constant which depends on the temperature and
not on the distance. In other words the running of the coupling at
finite temperature is controlled by the distance $r$ for small $r$ and
by the temperature for $r \gg 1/T$. 
\begin{figure}
\includegraphics[width=2.6in]{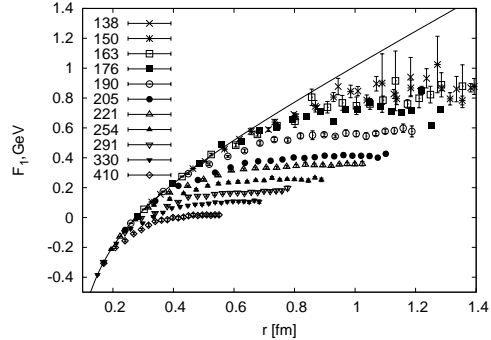}
\vspace*{-1cm}
\caption{
The singlet free energy in three flavor QCD \cite{petrov04}.
The temperature is given in $MeV$.}
\label{freee_nf3}
\vspace*{-0.8cm}
\end{figure}
The results presented so far refer to $SU(3)$ gauge theory.
Very recently the study of the static $Q \bar Q$ free energies 
was extended to two and three flavor QCD 
\cite{kaczmarek03dyn,petrov04,toth04}. 
The results for the singlet free energy in 
3 flavor QCD are shown in Fig. \ref{freee_nf3}. 
As in the case of pure gauge theory
at short distances the free energy is temperature independent and coincides
with the zero temperature potential. However, the free energy reaches a 
constant value also at low temperatures, it flattens at distances of 
about $0.9fm$. This is usually interpreted as string breaking.

The free energies of static quark anti-quark pairs are often called
and interpreted as potentials. They have been used in potential
models to predict the pattern on $J/\psi$ suppression \cite{digal01}. 
However, one should be aware that in fact one deals with free
energies which contain an entropy contribution. 
Having the partition function in the presence of static
$Q \bar Q$ pairs one can also calculate the entropy and the internal
energy difference between the system with and without static charges
\begin{equation}
V_1(r,T)=-T^2 \frac{\partial F_1(r,T)/T}{\partial T},
\end{equation}
\begin{equation}
S_1(r,T)=-T \frac{\partial F_1(r,T)}{\partial T}.
\end{equation}

\begin{figure}
\includegraphics[width=2.4in]{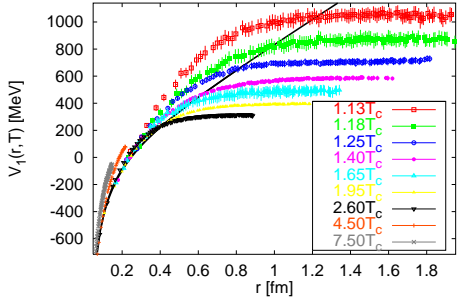}
\vspace*{-1cm}
\caption{The singlet internal energy above deconfinement 
in SU(3) gauge theory \cite{kaczmarek03lat}.}
\label{vinf}
\vspace*{-0.5cm}
\end{figure}

\begin{figure}
\includegraphics[width=2.3in]{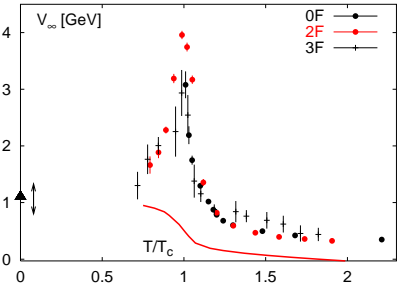}
\vspace*{-1cm}
\caption{The asymptotic value of the internal energy
energy  in SU(3) gauge theory \cite{kaczmarek03lat}, 
2 flavor QCD \cite{kaczmarek03dyn} and 3 flavor QCD \cite{petrov04}.
The line is the asymptotic value of the free energy in 2 flavor QCD.}
\vspace*{-0.5cm}
\label{v1}
\end{figure}

The free energy is decreasing with temperature
(see e.g. Fig. \ref{freee_nf3}). This implies that the entropy 
contribution is negative (c.f. equation above).
In Fig. \ref{v1} we show the internal energy at different temperatures.
It is obvious that the internal energy is larger than the
free energy (compare with Fig. \ref{free_all} ). If one would interpret the
internal energy as potential it would lead to the conclusion
that the $J/\psi$ remains bound till $1.7T_c$ \cite{shuryak04,wong04}. 
One may wonder to which extent such interpretation is correct.
In Fig. \ref{vinf} we show the asymptotic value of the internal energy
as function of the temperature. One can see a large increase of 
$U_{\infty}(T)$ in the vicinity of the transition. Such large
increase is difficult to interpret in terms of modification of
inter-quark forces and it is probably due to many body effects.

\section{Meson spectral functions}
Lattice QCD can provide information about imaginary time
correlation functions. 
On the other hand to make contact with
observable quantities (e.g. masses and widths of hadron
resonances in the medium) one should deal with correlators in real time
\begin{equation}
D^{<}(t)=\langle \hat O(0) \hat O(t) \rangle,~
D^{>}(t)=\langle \hat O(t) \hat O(0) \rangle,
\end{equation}
or equivalently the spectral function
\begin{equation}
\sigma(\omega)=\frac{D^{>}(\omega)-D^{<}(\omega)}{ 2 \pi}.
\end{equation}
For example the rate of emission of thermal dileptons can be 
expressed in terms of the spectral function of the correlator
of the vector current $J_{\mu}=\bar q \gamma_{\mu} q$ \cite{braaten90}.
\begin{equation}
{dW \over dp_0 d^3p}  = {5 \alpha_{em}^2 \over 27 \pi^2} 
{1 \over p_0^2 (e^{p_0/T}-1)} \sigma_V(p_0, \vec{p}).
\end{equation}

Here we consider correlators of point meson operators
$\bar q \Gamma q$, with 
$\Gamma=1,~\gamma_5,~\gamma_{\mu}, \gamma_{\mu}\gamma_5$ for 
scalar, pseudo-scalar, vector and axial-vector channels.
We will also restrict the discussion to the case of zero 
spatial momentum.
The imaginary time correlator is an analytic continuation of
the real time correlator $G(\tau)=D^{>}(-i \tau)$. Using this
and periodicity in the imaginary time (also known as 
Kubo-Martin-Schwinger condition) an integral representation for
$G(\tau)$ in terms of the spectral function can be derived
\begin{eqnarray}
G(\tau, T) &=& \int_0^{\infty} d \omega
\sigma(\omega,T) K(\tau,\omega,T) \label{eq.spect} \\
K(\tau,\omega) &=& \frac{\cosh(\omega(\tau-1/2
T))}{\sinh(\omega/2 T)}.
\label{eq.kernel}
\end{eqnarray}
This relation is valid in the continuum. It is not obvious that it
holds also for the correlators calculated on the lattice. 
However, it was shown that an integral representation like
in Eq. \ref{eq.kernel} can be also defined for
the lattice correlator in the limit of the free field 
theory \cite{karsch03fspf}.
The only modification is that the upper integration limit becomes
finite $\omega_{max} \simeq 4 a^{-1}$ \cite{karsch03fspf}. 
The cutoff dependence 
of the correlators is contained in the spectral function \cite{karsch03fspf}.
The obvious
problem is that we have to reconstruct several hundred degrees of freedom
(needed for a reasonable discretization of the integral 
in Eq. (\ref{eq.kernel})).
with ${\cal O}(10)$ data point on $G(\tau)$. This ill-posed problem can be 
solved using the Maximum Entropy Method (MEM) \cite{nakahara99}. 
The method has been applied to study meson spectral functions
at zero \cite{nakahara99,yamazaki02,blum04} 
as well as finite temperature 
\cite{karsch02dil,karsch03qm,asakawa03qm,petreczky03sqm,umeda02,asakawa03hq,datta04}.

The studies of meson spectral functions at zero temperature 
reproduced the properties of ground state mesons but on the 
other hand revealed structures (lattice artifacts) whose 
position scales like $a^{-1}$ \cite{yamazaki02,datta04,blum04}. 
The cutoff effects in the
meson spectral functions extracted from the 
interacting theory have little to do with those in the free theory.
\begin{figure*}
\includegraphics[width=2.5in]{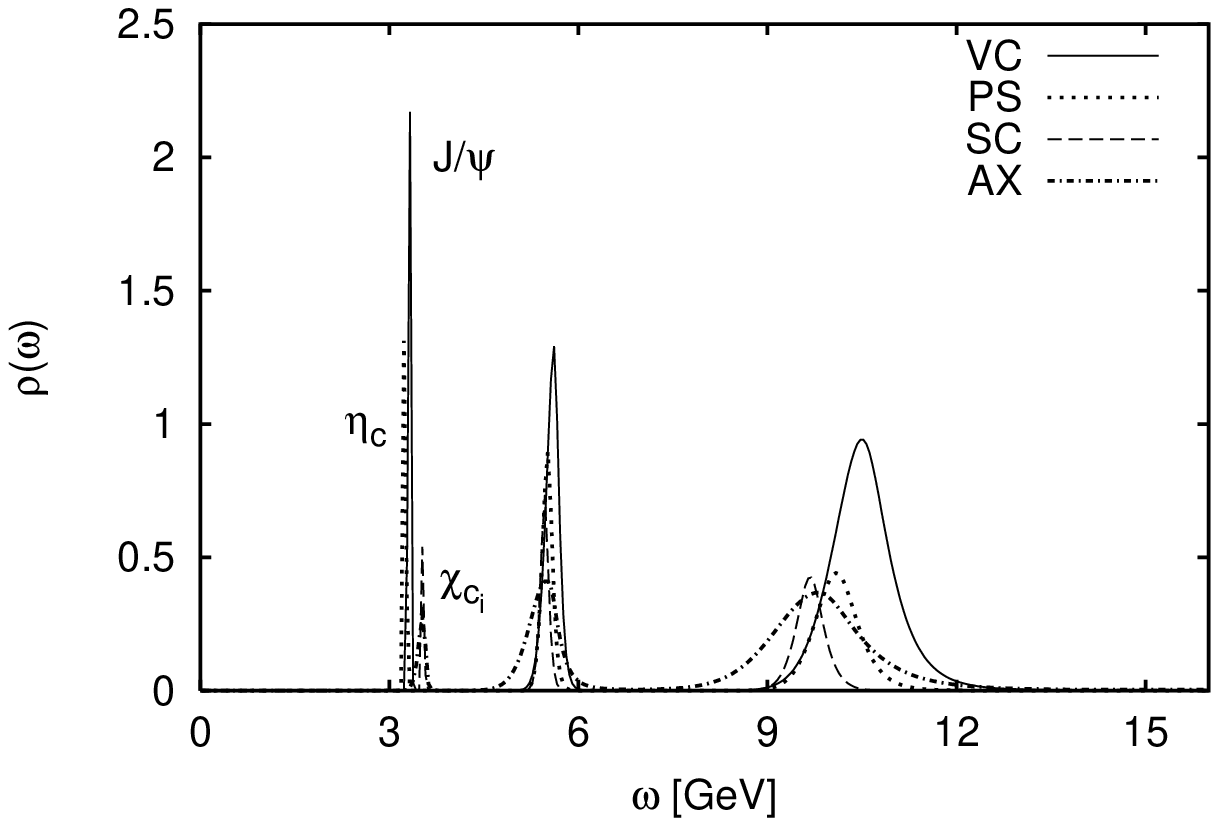}
\includegraphics[width=2.5in]{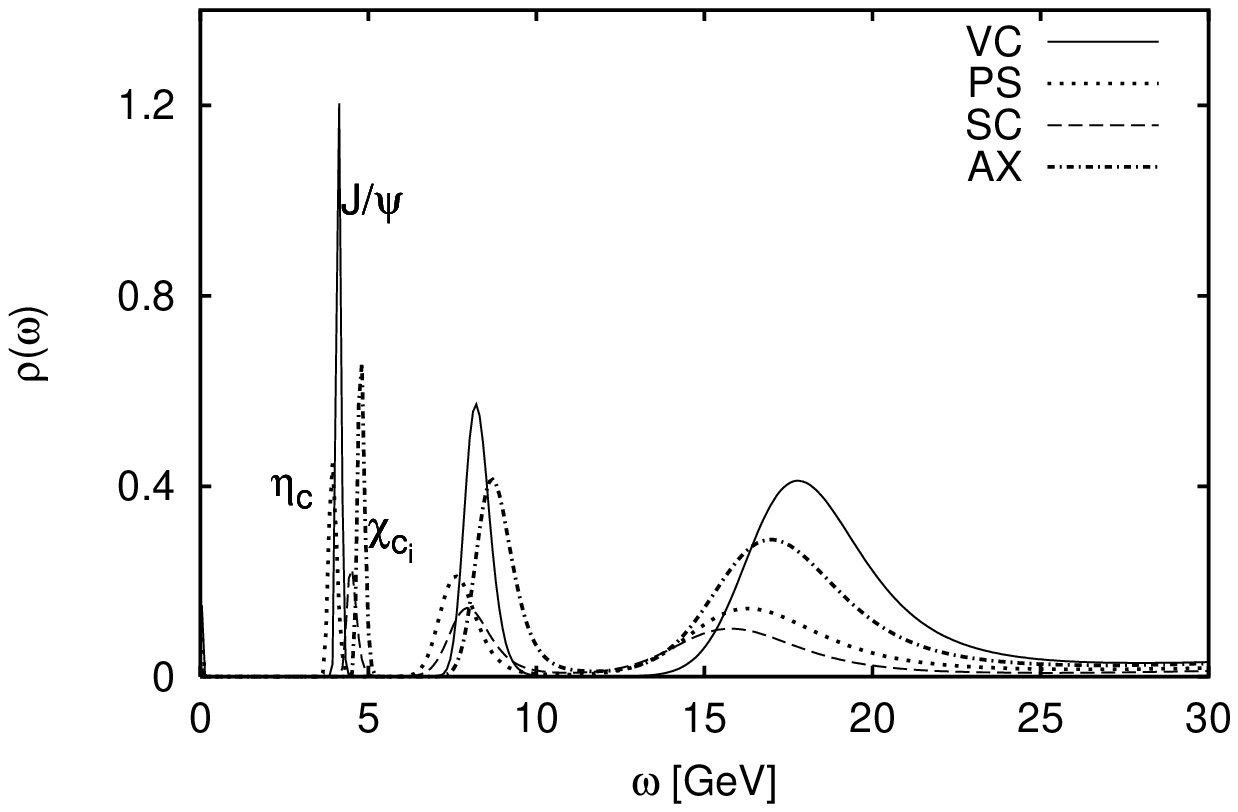}
\vspace*{-1cm}
\caption{The charmonia spectral functions below deconfinement
for two different lattice spacings $a^{-1}=4.86GeV$ (left)
and $a^{-1}=9.72GeV$ (right) \cite{datta04}.}
\label{comp_a_spf}
\vspace*{-0.5cm}
\end{figure*}

First I will discuss finite temperature meson spectral functions for 
heavy quarks. This is of particular interest as it was suggested by Matsui
and Satz that existence or non-existence of heavy quarkonia in the plasma
reflects the screening properties of the plasma and can be used
as signal for deconfinement and quark gluon plasma formation \cite{MS86}.
Different calculations based on screening and potential models have
estimated that $J/\psi$ will dissolve around $1.1T_c$ \cite{digal01,karsch88}.
Calculations
of quakonium spectral function using MEM were performed only
recently. Calculations have been done using anisotropic as well
as fine isotropic lattices in quenched approximation 
\cite{umeda02,asakawa03hq,datta04}. 
Whatever approach is used several spatial
lattice spacings are needed to check the cutoff dependence of the
result. To demonstrate this in Fig. \ref{comp_a_spf} we show
charmonium spectral functions below deconfinement at two different
lattice spacings from Ref. \cite{datta04}. 
The pseudo-scalar and vector channels
correspond to $\eta_c$ and $J/\psi$ states, while the scalar
and axial-vector channels correspond to $\chi_{c0}$ and
$\chi_{c1}$ states. One can see that while the ground state 
peaks reproduce the expected masses the positions
of the other peaks scales roughly as $1/a$, i.e. these peaks 
are lattice artifacts. Thus point operators can provide information 
only about the properties of the ground state for a given channel.

\begin{figure*}
\includegraphics[width=2.2in]{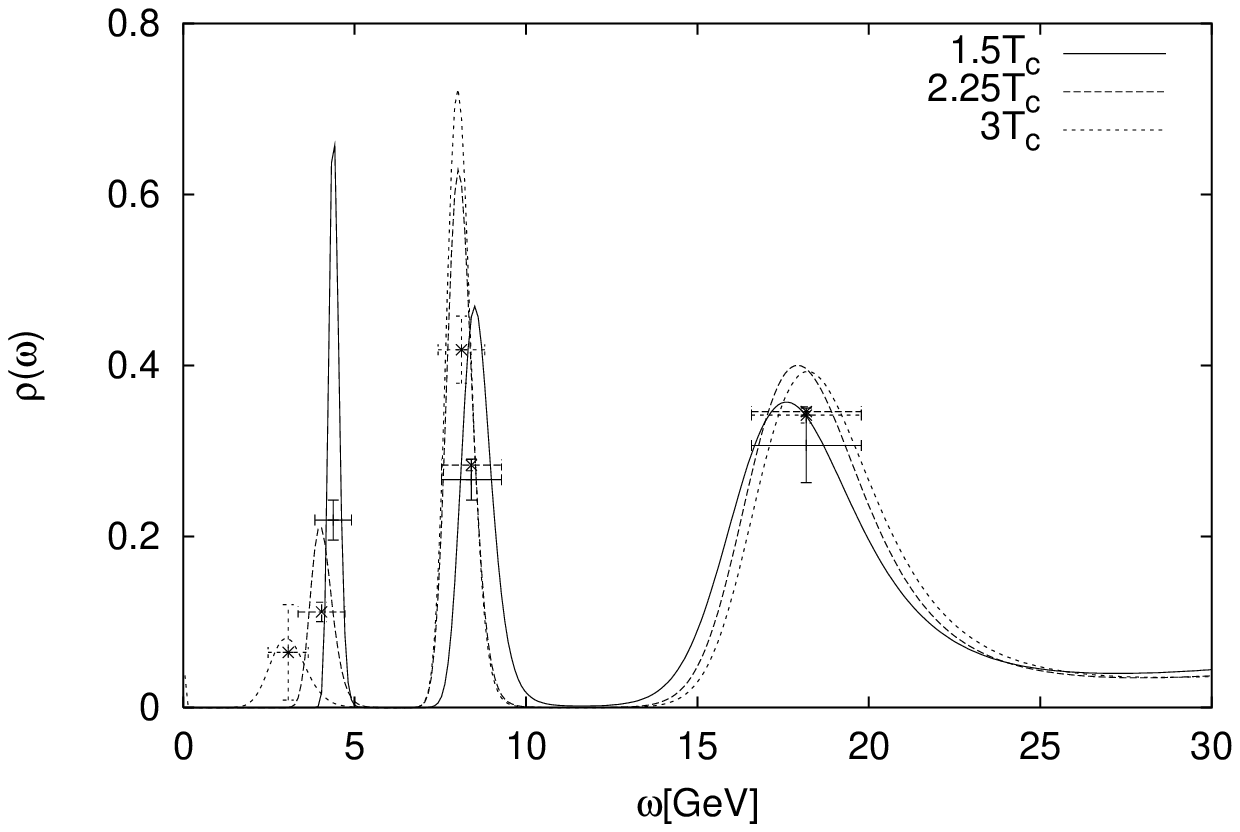}
\includegraphics[width=2.3in]{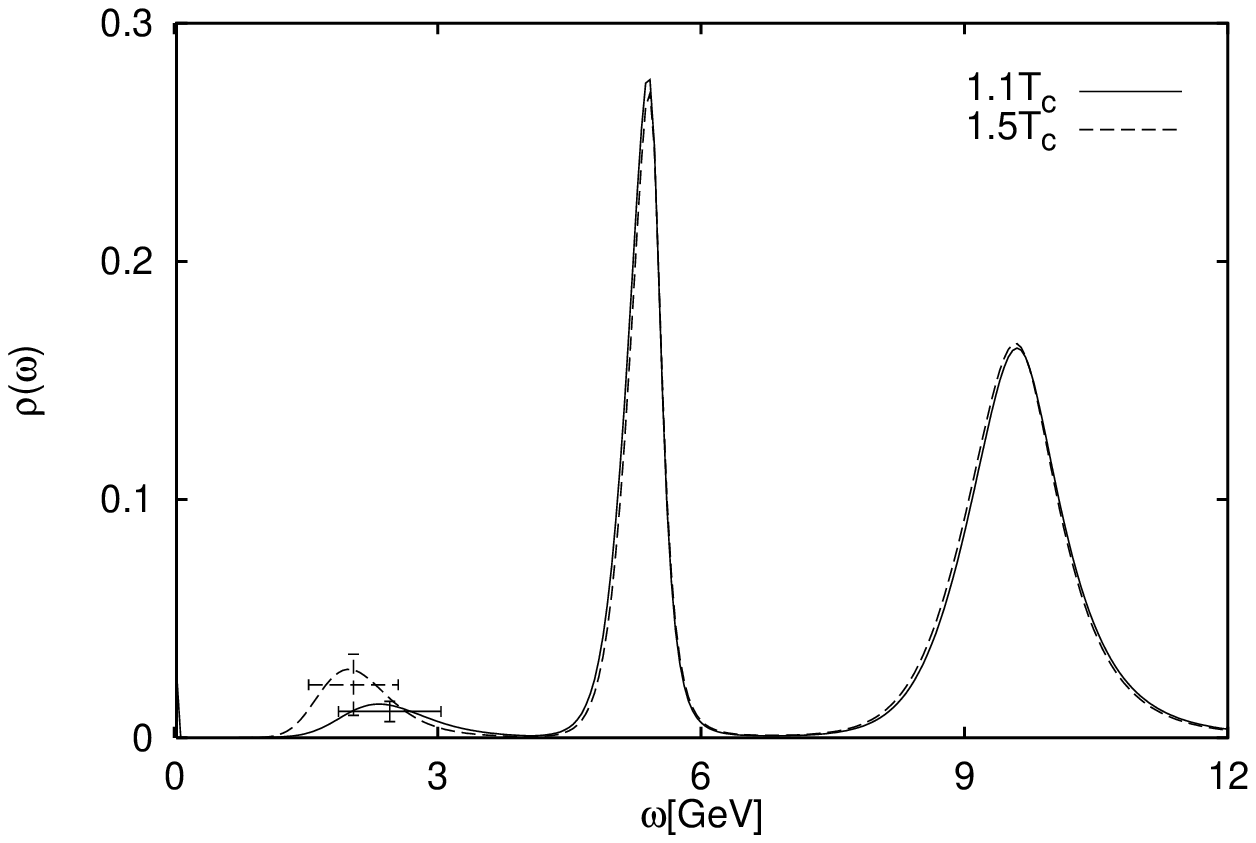}
\vspace*{-1.2cm}
\caption{Charmonia spectral functions for the vector (left) and 
scalar (right)
channels above $T_c$.
\cite{datta04}.}
\label{charm_spf_FT}
\vspace*{-0.6cm}
\end{figure*}

Now let me discuss what happens at finite temperature.  In this case
the reconstruction of the spectral function becomes more difficult
as not only the number of available time-slices is reduced but also
the extent of the imaginary time direction may be quite small
$<1/T$ (see discussion in Ref. \cite{umeda02} on this point).
While the former limitation can be relaxed using larger
anisotropies the later will always remain. Therefore it is useful
to inspect the temperature dependence of the correlators at the first
step. The temperature dependence of the correlators is due to
the explicit temperature dependence of the corresponding spectral functions 
and the temperature dependence of the integration kernel
in Eq. (\ref{eq.kernel}). To factor out the trivial temperature
dependence of the correlator due to periodicity in the time direction we 
construct the following model correlator
\begin{equation}
G_{rec}(\tau,T)=\int_0^{\infty} d \omega \sigma(\omega,T^{*})K(\tau,\omega,T),
\end{equation}
\begin{figure*}
\includegraphics[width=2.5in]{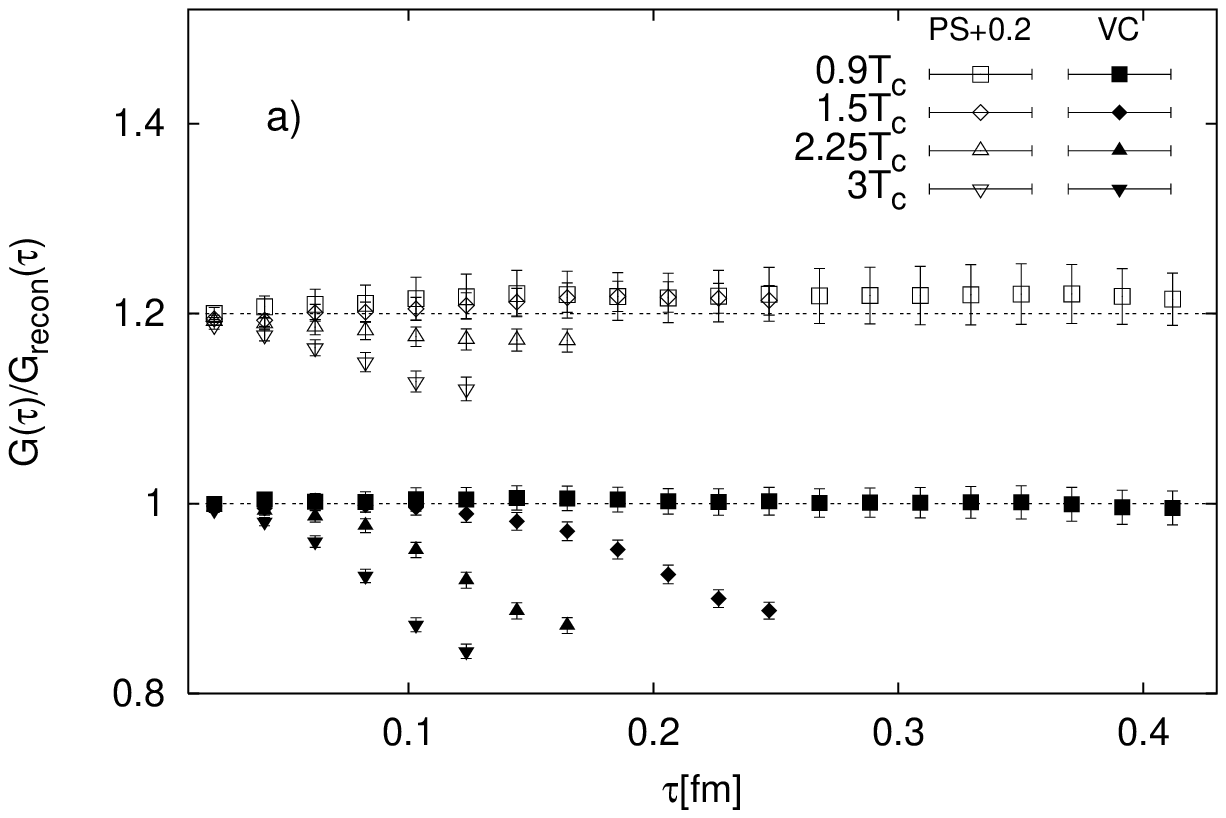}
\includegraphics[width=2.5in]{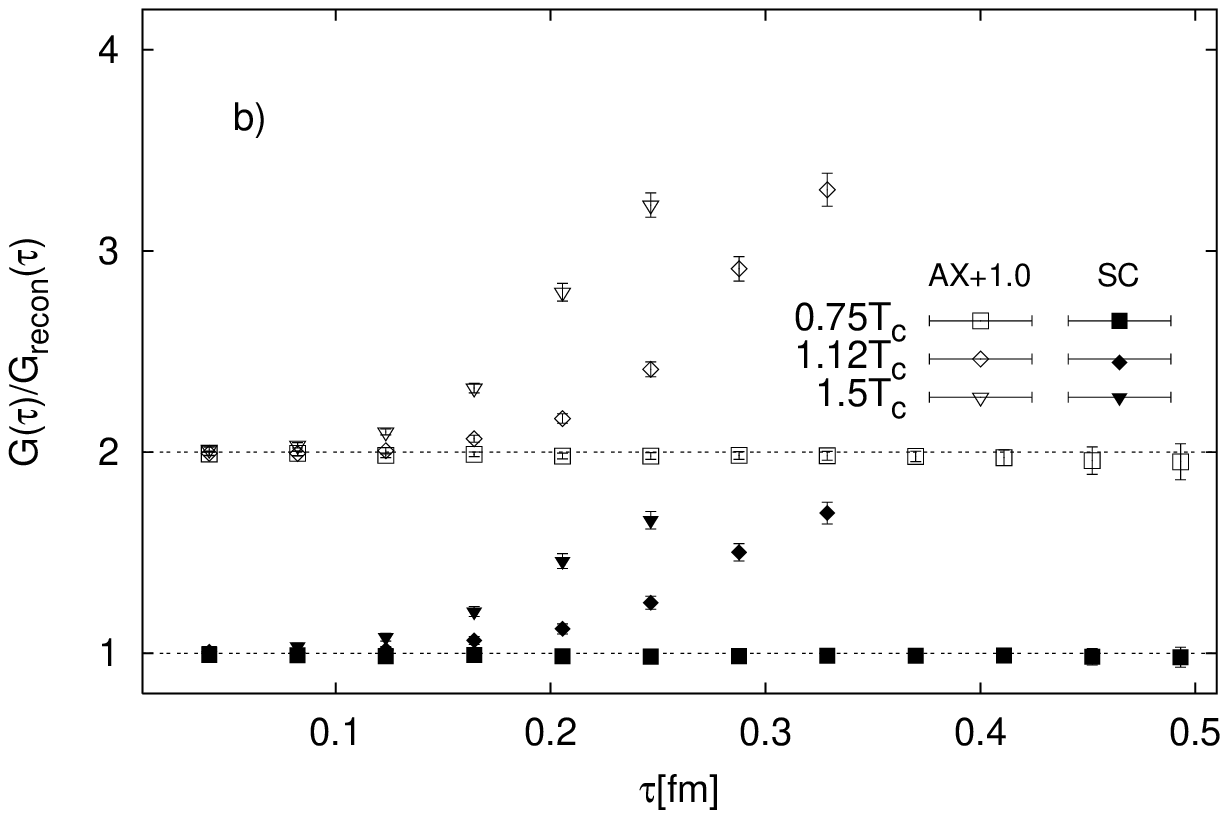}
\vspace*{-1.2cm}
\caption{Ratios $G/G_{recon}$ for the pseudo-scalar and the vector correlators
(left) and for the scalar and the 
axial-vector correlators (right) \cite{datta04}.
For better visibility in the pseudo-scalar and axial-vector channel the data
for $G/G_{recon}$ have been shifted by a constant.
}
\label{ratio}
\vspace*{-0.4cm}
\end{figure*}
with $T^{*}$ being some temperature below deconfinement. If the
spectral function has no temperature dependence also above deconfinement
we would expect no temperature dependence in $G/G_{rec}$
and $G/G_{rec} \simeq 1$.
In Fig. \ref{ratio} we show $G/G_{rec}$ for different channels. As one can see
for temperatures $T<2 T_c$ this ratio is not very far from unity for
the pseudo-scalar and vector channels, while for the scalar and axial-vector
channels large changes are visible already at $1.1T_c$. This suggests
that ground state charmonia may survive deconfinement with little 
modifications of their properties, while excited charmonia states
are dissolved or strongly modified by the medium right after 
deconfinement. To study the problem more in detail one
needs to reconstruct the spectral functions which are shown in 
Fig. \ref{charm_spf_FT}. 
The figure confirms the expectation based on the analysis of
the correlator, $J/\psi$ seems to exist up to temperatures
$2.25T_c$ , while the $\chi_{c0}$ states is dissolved at $1.1T_c$.
The situation is similar for the pseudo-scalar and axial vector
spectral functions. The results for the $1S$ states ($J/\psi$, $\eta_c$)
are confirmed by calculations done on anisotropic lattices, though
there is some controversy about the temperature where $J/\psi$ dissolves
and the way it disappears. In Ref. \cite{asakawa03hq} 
it was found that the $1S$ charmonia
abruptly disappear at $1.7T_c$, while the spectral functions shown above
suggest a gradual dissolution.

\begin{figure}
\includegraphics[width=2.2in]{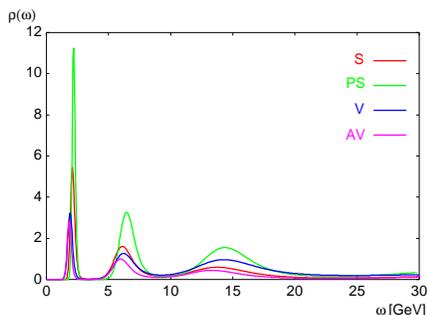}
\vspace*{-1cm}
\caption{Spectral functions above deconfinement for
$m_{PS}/m_{V}=0.65$ \cite{asakawa03qm}, i.e. for
quark masses around the strange quark mass.}
\vspace*{-0.7cm}
\label{light_spf}
\end{figure}

Meson spectral functions at finite temperature were also studied in the
case of light quarks. In particular in Ref. \cite{karsch02dil} the first attempt to calculate
vector spectral functions and the thermal dilepton rate was made using
isotropic lattices. Then meson spectral functions for scalar, 
pseudo-scalar, vector and axial-vector channels were calculated 
using anisotropic lattices \cite{asakawa03hq}. 
One should keep in mind, however, that
unlike in heavy quark case here no study of possible 
systematic effects was done
so far. One common expectation that the pseudo-scalar and scalar
correlators should become degenerate at high temperatures due to
effective restoration of $U_A(1)$ symmetry. The same is true 
for the vector and axial-vector correlators. 
Explicit calculations confirm these expectations 
\cite{karsch03qm,petreczky03sqm}.
An interesting feature of the light meson spectral function above
deconfinement is that they do not show any continuum like structures
which one could identify with freely propagating quarks but peaks.
In Fig. \ref{light_spf}  I show the meson spectral functions at $1.4T_c$ which
demonstrates this feature. While the second and third peak are 
likely to be lattice artifacts (though not yet shown by a  detailed analysis)
the first peak is likely to be physical and its existence is quite bizarre.
Note that the spectral functions in Fig. \ref{light_spf} 
also show the degeneracy of
opposite parity channels as expected from the analysis of the correlators, 
moreover the masses in all four channels appear
to be roughly the same.
\vspace*{-0.3cm}
\section*{Acknowledgment}
\noindent
This work has been authored under the contract 
DE-AC02-98CH10886 with the U.S. Department of energy.
I would like to thank Z. Fodor and F. Karsch for careful
reading of the manuscript and many valuable comments.

\end{document}